\documentstyle{article}
\title{\bf Quantum Gravity and the Problem of Measurement}
\author{ Pedro F. Gonz\'alez-D\'{\i}az$^{1}$ \\
$^{1}$Instituto de Matem\'aticas y F\'{\i}sica Fundamental\\
Consejo Superior de Investigaciones Cient\'{\i}ficas\\
Serrano 121, 28006 Madrid (Spain) }
\date{}

\begin{document}

\maketitle
\large
\setlength{\baselineskip}{0.5cm}

\vspace{1.5cm}

\noindent {\bf Running Head}: Quantum Gravity and Measurement

\pagebreak

\noindent{\bf Abstract.}
We discuss some arguments in favour of the proposal that the quantum correlations
contained in the pure state-vector evolving according to Schr\"odinger equation
can be eliminated by the action of multiply connected wormholes during
measurement. We devise a procedure to
obtain a proper master equation which governes the changes of the reduced density
matrix of matter fields interacting with doubly connected wormholes. It is shown
that this master equation predicts an appropriate damping of the off-diagonal
correlations contained in the state vector.

\vspace{1.5cm}

\noindent {\bf Key words}: wormholes, measurement, mixed states, master equation

\pagebreak

\section{\bf The proposal}

There have been some proposals to provide the necessary decay of quantum coherence
during measurement with a physical explanation (Zurek 1991).
The most popular among these proposals
is the so-called decoherence program advocated mainly by Zeh 1970,
Gell-Mann and Hartle 1990
on the one hand, and by Unruh and Zurek 1989, on the other.
It is not quite clear
however that these proposals could provide with a consistent mechanism to generate
the causal nonlocality and arrow of time which are also induced by quantum
measurement (Bell 1975, Omn\`es 1994).
Another approach to the quantum measurement problem has been advocated by
Penrose 1987. It adscribes the cause leading to the wave packet collapse to some
still unspecified form of gravitational interaction.
In this paper we shall consider a nonunitary evolution of microscopic
systems arising from a violation of causal
locality which can only be induced by {\it nonsimply} connected wormholes
(Gonzalez-D\'{\i}az 1992).

Wormholes are microscopic connections
between two otherwise desconnected flat regions of spacetime and represent
a topology change in that they induce an initial state which is just flat space
to evolve into a final state which is flat space plus a given number of baby
universes (Hawking 1990).
We distinguish two possible inner topologies
which may appear in wormholes. If the inner topology is simply connected, the corresponding
quantum state is a pure state described by a wave function (Hawking 1990).
However, if the inner topology of the wormholes
is not simply connected, the quantum state becomes mixed and should be
described by a
density matrix (Gonzalez-D\'{\i}az 1991).

The interaction
formalism worked out by using the technique of the Green function filtered by
the wormhole quantum state (Gonzalez-D\'{\i}az 1991, 1993)
gives rise to a bilocal interaction contribution
$P_{i}I_{bl}(x_{1},x_{2};y_{1},y_{2})$, where the bilocal factor
$I_{bl}$ is independent of the
wormhole state. The prefactor $P_{i}$ depends however on that state. If we start
with a wormhole wave function, $P_{i}\equiv P_{\Psi}$ is just an unimportant
numerical coefficient of order unity, but if we choose a density matrix as the
wormhole state, then $P_{i}\equiv P_{\rho_{b}}(n_{j})$ will depend on the wormhole
energy spectrum (Gonzalez-D\'{\i}az 1993, 1994, 1995).
The quantum state of wormholes is generally given by a path integral
\begin{equation}
F_{w} = \int_{C} d[g_{\mu\nu}]d[\Phi_{0}]e^{-I[g_{\mu\nu},\Phi]},
\end{equation}
in which $I$ is the Euclidean action and $C$ represents the class of asymptotically
flat Euclidean four-geometries and asymptotically vanishing matter field configurations
which match either the prescribed data on a three-surface dividing the four-manifold
in the case of a pure state given by a wave function $\Psi$
(Hawking 1988), or the data on its
three-surface and the orientation reverse of the corresponding set of data on its
copy three-surface in the case of a mixed state given by a
density matrix $\rho_b$ (Gonz맓ez-D며z 1991).
If $F_{w}=\Psi$, one can apply the Gell-Mann-Low formula and obtain an effective
interaction Hamiltonian given by the Hawking-Coleman expression
(Hawking 1988, Coleman 1988),
$\sum_{k} H_{k}^{I}(\Phi)(a_{k}^{\dagger}+a_{k})$,
where the discrete index $k$ collectively labels the types of different baby universes,
the $H_{k}^{I}$'s are matter-field interaction Hamiltonian terms, and the $a_{k}$'s
are Fock operators for the baby universes. This
leads (Hawking 1990) to no loss of quantum coherence and implies (Gonzalez-D\'{\i}az 1994)
that the quantum state of a simply connected wormhole contains
an equal contribution from complex and their complex conjugate metrics, so that
causal locality (i.e. $[H^{I}(x),
H^{I}(y)]=0$) holds (Hawking 1990) and both $\Psi$ and the quantum state of the
matter field should then be time-symmetric. On the contrary,
if $F_{\omega}=\rho_{b}$, one cannot apply Gell-Mann-Low formula. Using then a combinatorial
procedure, we obtain (Gonzalez-D\'{\i}az 1992):
$\sum_{k}H_{k}^{I}(\Phi)A_{k}(a_{k}^{\dagger},a_{k})$,
where $A_{k}$ generally contains higher, nonlinear powers of the baby-universe
Fock operators. Restricting to the simplest case in which the inner wormhole
topology is doubly connected (Gonzalez-D\'{\i}az 1992),
$A_{k} \simeq a_{k}^{+2} + a_{k}^{2}$ and
\begin{equation}
\int d^{4}x_{0}[H^{I}(x),H^{I}(x')] = D_{0}(c^{\dagger}c+\frac{1}{2})\sinh(2k_{0}),
\end{equation}
where $D_{0}$ is a constant of order unity,
with the $c$'s being Fock operators for the matter field, and
$k_{0} = (\frac{2k}{R_{0}^{2}-k})^{1/2}$,
in which $(2k)^{1/2}$ denotes the proper separation distance, measured
on the wormhole inner three-manifold, between the two correlated points at which two
baby universes are created or annihilated, the two at a time, and
$(R_{0}^{2}-k)^{1/2}$ gives the length scale of the baby universes. Then
the demand of Lorentz invariance implies (Gonzalez-D\'{\i}az 1992, 1994) that (1.1) leads to
loss of quantum coherence so as a breakdown of time symmetry and causal
locality, at least for $CP$ invariant matter with positive energy. Thus,
nonsimply connected wormhole fluctuations are able to induce all the
effects which are required for quantum measurements.
We summarize then our proposal as follows.
The unitary evolution of the state vector
governed by Schr\"odinger equation should be associated with a physical system
interacting with wormholes which are all
in a pure state, while the nonunitary
quantum-measurement evolution should correspond to a physical
situation in which the system interacts with at least a nonzero
proportion of wormholes in mixed state.

Our proposal needs to be implemented, however, in two important respects. First
of all, it is not still quite clear whether the loss of quantum coherence induced
by the breakdown of casual locality implied by (1.2) is of the kind required by
quantum measurement to damp the off-diagonal correlations in the state vector in
a sufficiently short decoherence time. This question will be addressed in the
next sections. Secondly,
one would need to provide the scenario with a reasonable mechanism by which
one could answer the question, what is the cause why one should suddenly replace
simply connected wormholes by multiply connected wormholes when some quantum
measurement is being carried out on a system?. We shall devote the rest of the
present section to briefly comment on this question.
Wormholes were initially claimed
(Coleman 1988) to fix the observed values of all physical and cosmological
constants. Actually, this can only be the case when multiply connected wormholes are
considered; otherwise, the topological fluctuations induce unphysical values for
these constants, such as either zero or infinite values for the physical constants or,
more importantly, a more probable large negative value for the cosmological
constant. It is only for statistical wormholes
that the whole set of observable effects induced by them on ordinary matter at low energies
can be thought to produce some anthropic consequences, along with observability
of the matter states and their couplings (Gonz맓ez-D며z 1993).
Indeed, out from the set of all possible
values of the physical and cosmological constants, such wormholes should take
on those values which will satisfy the requirements that there exist sites in the
universe where carbon-based life can evolve, thus rendering any form of anthropic
principle (Barrow and Tipler 1986)
just a more effect induced by statistical wormholes on matter.
On the other hand, it has been shown (Gonz맓ez-D며z 1994)
that it is just the purely statistical
content of single mixed wormhole states which can give rise to the emergence of a
consistently defined cosmological time concept. There is no similar time concept
arising from wormholes when they are in a pure state. The above question becomes then meaningless, since asking about anything prior
to the emergence of time does not make any sense. Moreover, the time asymmetry
which is also induced by single wormhole statistical states can be regarded as
the common physical origin for all existing time arrows, including the so called
phychological arrow by which observers are able to remember just those records
in their memory produced by measurements already performed.
It appears then that there are some strong link-if not direct correspondence-
between the appearance of nonsimply connected wormholes and both, the existence
of observers able to perceive the flow of time and record the results of measurements,
and the simultaneous emergence of a causally-connected $classical$ reality endowed
with a set of observables.

\section{\bf The Master Equation}
\setcounter{equation}{0}

We shall restrict to the simplest case of interaction between
a massless, conformally-invariant scalar matter field, $\Phi(x,t)$, and
single doubly connected wormholes (Gonzalez-D\'{\i}az 1992), and use the formalism of the
density matrix in the interaction representation and ordinary time-dependent
perturbation theory in first-order approximation.
We start with an interaction Hamiltonian
$H^{I}= \sum_{i}H_{i}^{I}(\Phi)A_{i}$
and assume a full density matrix
$\rho = \rho_{\Phi} \otimes \rho_{b} = \rho_{\Phi}\sum_{i,j} |i><j|$,
where $\rho_{\Phi}$ and $\rho_{b}$ are density matrices for the scalar matter field
and the baby universes, respectively. From the equation of motion for $\rho$, which
we iterate for small increment of time, we obtain with the same
approximation as used in ordinary time-dependent perturbation calculations,
\begin{equation}
\dot{\rho}_{\Phi}(k,t) = Tr_{b}\sum_{i,j}[H_{i}^{I}A_{i},[H_{j}^{I}A_{j},\rho]],
\end{equation}
in which $Tr_{b}$ means tracing over the baby universe operators.

Also, we assume an orthonormalization relation
\begin{equation}
<j|i> = \delta(2k-(x-x')^{2})\delta(p^{2}-(R_{0}^{2}-k)^{-1}),
\end{equation}
and commutation relations for hermitian operators $A_{i}$
(Gonz맓ez-D며z 1992)
\begin{equation}
\int d^{4}x_{0}[A(x),A(x')] = E_{0}(a^{\dagger}a+\frac{1}{2})\sinh[\frac{8k}{R_{0}^{2}-k}]^{1/2},
\end{equation}
where $x_{0}=\frac{1}{2}(x-x')$ and $E_{0}$ is a constant c-number of order unity.
The baby universe commutator (2.3) will consistently vanish at the limit $k \rightarrow 0$.

Using then the usual Fock expansion for quantum-field operators, and introducing
the substitution (Gonzalez-D\'{\i}az 1992) $\sum_{i,j} \rightarrow \int d^{4}x_{0}$, we obtain, after
integrating over $x_{0}=\frac{1}{2}(x-x')$ and momentum $p$, with the customary
measure $d\tilde{P}=\frac{dp}{(2\pi)^{3}}\delta(p^{2}-(R_{0}^{2}-k)^{-1})\theta(p_{0})$,
the master equation for the reduced density matrix $\rho_{\Phi}$ of the scalar
matter field $\Phi$:
\[
\dot{\rho}_{\Phi}(k_{0},t)=A(k_{0},N)O_{4}(k_{0},c)\rho_{\Phi} + B(k_{0},N)\rho_{\Phi}O_{4}(k_{0},c)
- 2[C(k_{0},N)c^{2}\rho_{\Phi}c^{\dagger 2} \]
\begin{equation}
+ D(k_{0},N)O_{4}^{\rho}(c) +
F(k_{0},N)c^{\dagger 2}\rho_{\Phi}c^{2}],
\end{equation}
where $N=0,2,4,...$ denotes the initial number of baby universes,
$k_{0}=(\frac{2k}{R_{0}^{2}-k})^{1/2}$, $Q(k_{0},N)= E_{0}e^{-2k_{0}}(2N+1)\sinh(2k_{0})$,
\begin{equation}
O_{4}(k_{0},c)=e^{4k_{0}}c^{2}c^{\dagger 2} + 4[(c^{\dagger}c)^{2}+c^{\dagger}c+\frac{1}{4}]e^{2k_{0}}+c^{\dagger 2}c^{2},
\end{equation}
\begin{equation}
O_{4}^{\rho}(c)=2(2c^{\dagger}c\rho_{\Phi}c^{\dagger}c + c^{\dagger}c\rho_{\Phi}+
\rho_{\Phi}c^{\dagger}c+\frac{1}{2}\rho_{\Phi}),
\end{equation}
\begin{equation}
A(k_{0},N)=(N+1)(N+2)e^{-4k_{0}}+N(N+1)
+\frac{1}{4}e^{-2k_{0}}+2Q(k_{0},N) ,
\end{equation}
\begin{equation}
B(k_{0},N)=A(k_{0},N)-2Q(k_{0},N),
\end{equation}
\begin{equation}
C(k_{0},N)=(N+1)(N+2)+[N(N+1)+\frac{1}{4}]\cosh(2k_{0})+Q(k_{0},N),
\end{equation}
\begin{equation}
D(k_{0},N)=2(N+1)^{2}cosh(2k_{0})+\frac{1}{4}+Q(k_{0},N),
\end{equation}
and
\begin{equation}
F(k_{0},N)=[(N+1)(N+2)+\frac{1}{4}]\cosh(2k_{0})
+N(N+1)+Q(k_{0},N).
\end{equation}

In order to eliminate changes in the reduced
density matrix $\rho_{\Phi}(k_{0},t)$ which do not originate from
quantum nonlocality,
we first
take the limit $k \rightarrow 0$ in (2.4) and substract then the
resulting expression (which is
associated with virtual processes by which scalar field quanta are created and
annihilated in such a way as to contribute the full master equation with terms
proportional to the Fock operator products $c^{2}c^{\dagger 2}$, $c^{\dagger 2}c^{2}$,
$c^{2}\rho_{\Phi}c^{\dagger 2}$ and $c^{\dagger 2}\rho_{\Phi}c^{2}$)
from (2.4) to finally obtain
a reduced density matrix for matrix elements in the Fock space of matter
field number states,
in diagonal representation
$\bar{P}_{n}(k,t)$, such that $\dot{\bar{P}}_{n}(0,t)=0$ results from a minimal
condition for the vanishing of the nonlocal effects in $\dot{\bar{P}}_{n} (k,t)$
at $k \rightarrow 0$, where
\begin{equation}
\dot{\bar{P}}_{n}(k,t)=-P(N,k_{0})(n+\frac{1}{2})^{2} \bar{P}_{n}(k,t),
\end{equation}
with
$P(N,k_{0})=8(N+\frac{1}{2})\sinh(2k_{0})$.
Eqn. (2.12) consistently
reduces to zero only when the parameter $k \rightarrow 0$. Note that even for
the vacuum states $N=n=0$, $\bar{P}_{n}(k,t)$ is not time invariant, but gives
rise to a residual zero-point loss of quantum coherence.

\section{\bf The Transition from Quantum to Classical}
\setcounter{equation}{0}
The density matrix $\bar{\rho}_{\Phi}$ in scalar particle number representation
$\bar{P}(k,t)$ corresponds actually to a quantum state in position representation.
The wormhole parameter $(2k)^{1/2}$ is not but the particular value of the
spacelike separation $(x-x')$ which coincides with the proper separation distance,
measured on a cross section of the inner wormhole manifold, between the two
correlated points at which baby universes are created or annihilated in
pairs. Hence, the evolution of the density matrix
$\bar{P}_{n}(x,x',t)$ will satisfy an accordingly generalized master equation,i.e.
\begin{equation}
\dot{\bar{P}}_{n}(x,x',t)=-P(N,x,x')(n+\frac{1}{2})^{2}\bar{P}(x,x',t),
\end{equation}
with
\begin{equation}
P(N,x,x')=8(N+\frac{1}{2})\sinh[\frac{4(x-x')^{2}}{r_{0}^{2}}]^{1/2},
\end{equation}
where $r_{0} = (R_{0}^{2}-k)^{1/2}$ is the radius of the throats
in the wormhole.

A coherent superposition
of two Gaussians separated by a distance equal to $\Delta x=x-x'$ will
now be considered (Zurek 1991).
At macroscopic distances, $\Delta x$ will be much larger than the Gaussian width,
so that the density matrix describing the state vector will have four peaks, two
at $x=x'$ (i.e. the diagonal elements which should survive the wave packet
collapse) and two at $x\neq x'$ (i.e. the off-diagonal elements
responsible for quantum correlations which
should disappear during the measurement process, giving rise to position as an
exactly preferred basis). Clearly, as required by quantum measurement,
Eqn. (3.1) will not produce any effects on the diagonal peaks, but will
induce a substantial damping of the off-diagonal peaks.
The parameter $N$ appearing in (3.1) is proportional to the initial population
of baby universes. Hence, in semiclassical aproximation, (3.2)
can be written
\begin{equation}
P(N,x,x') \equiv P(S_{\omega},x,x')=8e^{-S_{\omega}}\sinh[\frac{4(x-x')^{2}}{r_{0}^{2}}]^{1/2},
\end{equation}
where $S_{\omega}$ is the Euclidean action of the wormhole.
For nonsimply connected wormholes, the path integral
which describes the effects of wormholes on ordinary matter is given in terms of
a Planckian probability (Gonzalez-D\'{\i}az 1993, 1995), $\pi(\alpha)$,
for the Coleman $\alpha$ parameters (Hawking 1990),
and hence the semiclassical nucleation rate for baby universes,
\begin{equation}
e^{-S_{\omega}} = \alpha^{2}[2ln(1+\pi(\alpha)^{-1})]^{-1},
\end{equation}
would appear to play the role of an equilibrium temperature, $T_{b}$, for our wormhole-
scalar particle system.

On the other hand, the factor $r_{0}^{-1}$ in the argument of the
hyperbolic sinus in (3.3) must be proportional to the mass,
$m$, acquired by the
scalar particles while interacting with the wormholes, and the whole
factor $P(T_{b},x,x')$ would be associated with the extreme case
where only wormholes which are nonsimply connected are involved
in the interaction. Since in quantum gravity there should also
exist a given, generally nonzero contribution from simply connected
wormholes, in order to account for this contribution
one should modify the whole expression (3.1) by introducing
an overall factor $0\leq\gamma_{b}\leq 1$, interpretable as a
rate of coherence loss induced by interaction
with wormholes, so that the full master equation would become
\begin{equation}
\dot{\bar{P}}_{n}(x,x',t) \sim -\gamma_{b} T_{b}\sinh[(x-x')m](n+\frac{1}{2})^{2}\bar{P}_{n}(x,x',t).
\end{equation}
Our master equation can now be compared
with the term responsible for Brownian
fluctuations (Zurek 1991) of the master equation in current
decoherence programs
\begin{equation}
\dot{\rho}_{CL}^{(f)} = -2M \gamma T (x-x')^{2} \rho_{CL}^{(f)},
\end{equation}
where the relaxation rate $\gamma=\frac{\eta}{2M}$, with $\eta$ being the
interaction viscosity, $T$ is the field temperature and $M$ the mass.
It is seen that although (3.5) and (3.6)
depend on the similar parameters in the same qualitative way,
(3.5) shows quite stronger a dependences on $(x-x')$ and mass
for large values of such parameters. Note furthermore that
(3.5) depends on the square of particle number.
In any case, these two expressions
damp the off-diagonal correlations in the qualitative way
required by quantum measurement.

\noindent\section*{Acknowledgements}
This work was supported by a CAICYT
Research Project N\mbox{$^{\underline{o}}$} PB94-0107.

\noindent\section*{References}
\begin{description}
\item [] J.D. Barrow and F.J. Tipler 1986, The Anthropic Cosmological
Principle (Clarendon Press, Oxford, UK)
\item [] J.S. Bell 1975 Helv. Phys. Acta 48, 93.
\item [] S. Coleman 1988 Nucl. Phys. B307, 867; S. Coleman 1988 Nucl. Phys. B310, 643.
\item [] M. Gell-Mann and J.B. Hartle 1990, in: Complexity, Entropy and the Physics
of Information, ed. W.H. Zurek (Addison-Wesley, Redwood City, California).
\item [] P.F. Gonz\'{a}lez-D\'{\i}az 1990 Phys. Rev. 42D, 3983.
\item [] P.F. Gonz\'{a}lez-D\'{\i}az 1991 Nucl. Phys. B351, 767.
\item [] P.F. Gonz\'{a}lez-D\'{\i}az 1992 Phys. Rev. D45, 499;
Phys. Lett. B293, 294.
\item [] P.F. Gonz\'{a}lez-D\'{\i}az 1993 Class. Quant. Grav. 10, 2505;
Mod. Phys. Lett. A8, 1089.
\item [] P.F Gonz\'{a}lez-D\'{\i}az 1994, in: Physical Origins of Time-Asymmetry, eds.
J.J. Halliwell, J. P\'erez-Mercader and W.H. Zurek (Cambridge Univ. Press, Cambridge);
Int. J. Mod. Phys. D3, 549.
\item [] P.F. Gonz\'{a}lez-D\'{\i}az 1995, in: Geometry of Constrained Dynamical
Systems, ed. J.M. Charap (cambridge Univ. Press, Cambridge).
\item [] S.W. Hawking 1988 Phys. Rev. D37, 904;
\item [] S.W. Hawking 1990 Mod. Phys. Lett A5, 145; 453.
\item [] R. Omn\`es 1994, in: Physical
Origins of Time-Asymmetry, eds. J.J. Halliwell, J. P\'erez-Mercader and
W.H. Zurek (Cambridge Univ. Press, Cambridge).
\item [] R. Penrose 1987, in: 300 Years of  Gravitation, eds. S.W. Hawking and W. Israel
(Cambridge Univ. Press, Cambridge)
\item [] W.G. Unruh and W.H. Zurek 1989 Phys. Rev. D40, 1071.
\item [] H.D. Zeh 1970 Found. Phys. 1, 69.
\item [] W.H. Zurek 1991 Phys. Today, pag. 36.

\end{description}

\end{document}